\documentclass[hidelinks,twocolumn, prl, superscriptaddress,a4paper]{revtex4}
\usepackage{graphicx}
\usepackage{dcolumn}
\usepackage{orcidlink}
\usepackage{csquotes}
\usepackage{float}
\usepackage{amsmath}
\usepackage{bm}
\usepackage{subfigure}

\begin{document}

\preprint{APS123-QED}

\title{Quantum chaos and complexity from string scattering amplitudes }

\author{Aranya Bhattacharya\,\orcidlink{0000-0002-1882-4177}}
\email{aranya.bhattacharya@th.if.uj.edu.pl}
\affiliation{Institute of Physics, Jagiellonian University, Łojasiewicza 11, 30-348 Kraków, Poland}

\author{Aneek Jana\,\orcidlink{0009-0001-1097-4250}}
\email{aneekjana@iisc.ac.in}
\affiliation{Centre for High Energy Physics, Indian Institute of Science,\\ C.V. Raman Avenue, Bangalore 560012, India.}

\date{\today}

\begin{abstract}
    
We introduce Krylov spread complexity in the context of black hole scattering by studying highly excited string states (HESS). Krylov complexity characterizes chaos by quantifying the spread of a state or operator under a known Hamiltonian. In contrast, quantum field theory often relies on S-matrices, where the Hamiltonian density becomes non-trivially time-dependent rendering the computations of complexity in Krylov basis exponentially hard. We define Krylov spread complexity for scattering amplitudes by analyzing the distribution of extrema, treating these as eigenvalues of a fictional Hamiltonian that evolves a thermo-field double state non-trivially. Our analysis of black hole scattering, through highly excited string states scattering into two or three tachyons, reveals that the Krylov complexity of these amplitudes mirrors the behavior of chaotic Hamiltonian evolution, with a pre-saturation peak indicating chaos. This formalism bridges the concepts of chaos in scattering and state evolution, offering a framework to distinguish different scattering processes.

\end{abstract}

\maketitle

\paragraph*{\textbf{Introduction.}---}

Complexity has emerged as a valuable tool for studying time-evolution dynamics in quantum systems and has garnered significant attention in holography for exploring spacetime structures beyond black hole horizons.\cite{Nielsen_2006, Jefferson:2017sdb, Parker:2018yvk, Barb_n_2019, Chapman:2017rqy, Balasubramanian:2022tpr, Brown_2016, Susskind:2014rva}. In general, complexity quantifies the relative difference between quantum states during Schrödinger evolution from an initial state. In the quantum circuit model, this is realized by counting the elementary quantum gates required to construct a target state from a reference state \cite{Nielsen_2006}. Recent studies have shown that complexity is optimally quantified in the Krylov basis, formed by the repeated action of the Hamiltonian \cite{Balasubramanian:2022tpr} on the initial state and measured by the average position of the evolved state in the Krylov basis. The notion of complexity in this context was first defined in the study of operator growth\cite{Parker:2018yvk}, and, has been particularly effective in probing quantum chaos and localization \cite{Parker:2018yvk, Balasubramanian:2022tpr, Rabinovici:2021qqt}.

While Krylov complexity has been extensively studied in diverse systems \cite{Rabinovici_2022, Caputa:2022eye, Afrasiar:2022efk, Balasubramanian:2022dnj, Caputa:2022yju, Erdmenger:2023shk, Pal:2023yik, Nandy:2023brt,Chattopadhyay:2023fob, Gautam:2023bcm, Bhattacharjee:2022qjw, Gill:2023umm, Bento:2023bjn, Aguilar-Gutierrez:2023nyk, Craps:2023ivc, Bhattacharya:2023yec, Caputa:2023vyr, Caputa:2024vrn, Huh:2023jxt, Nandy:2024htc,  Nizami_2023, nizami2024spreadcomplexityquantumchaos, Bhattacharya:2022gbz, Bhattacharjee:2022lzy, Bhattacharya:2023zqt, Bhattacharjee:2023uwx, Bhattacharya:2024uxx, Bhattacharya:2024hto, Dixit:2024pcj, Nandy:2024wwv, Amore:2024ihm, Bhattacharjee:2024yxj, Chapman:2024pdw, Balasubramanian:2024ghv, Anegawa:2024yia, Baggioli:2024wbz, Jha:2024nbl, Seetharaman:2024ket, takahashi2024krylovsubspacemethodsquantum, alishahiha2024krylovcomplexityprobechaos}, the formalism relies on the existence of a time-independent Hamiltonian to generate time evolution. However, in quantum field theories (QFT), the Hamiltonian density for quantum fields is usually always time-dependent. Moreover, in computing the $S$-matrix, which is the most natural observable in QFTs, the states are usually considered at far past ($-\infty$) and far future ($+\infty$) to get asymptotically free states. While there have been some efforts in studying chaos in QFTs using Krylov methods from thermal two point functions\cite{Avdoshkin:2022xuw, Camargo:2022rnt}, there is not yet a definite way to distinguish chaotic QFTs from non-chaotic ones, and extend the understanding to infinite timescales to understand $S$-matrices. In unrelated efforts, there have been description of chaos from many particle scattering amplitudes \cite{Rosenhaus:2020tmv, Gross:2021gsj, Rosenhaus:2021xhm} indicated by the erratic behavior of scattering amplitudes with respect to the scattering angle. In this case, there is no explicit Hamiltonian and hence no problem with time-dependence to be dealt with. However, one is yet to make the connection between these two existing notions of chaos, namely the Krylov complexity and the erratic amplitudes. We address this problem of defining Krylov basis for extremely time-dependent Hamiltonians in QFTs by leveraging these recent advances in understanding chaotic scattering amplitudes through the study of the position of their extrema \cite{Bianchi:2022mhs, Bianchi:2023uby, Bianchi:2024fsi}. We write down a fictional Hamiltonian from these positions of extrema and define a notion of complexity with the following broad motivations and generalizations in mind as listed below. 

\begin{itemize}
    \item We want to quantify the erratic behavior in scattering amplitudes through complexity in the Krylov basis. Our approach can be thought of as a way of repackaging the time-dependent QFT Hamiltonian interactions for times past to future infinity, from the scattering amplitudes, through a fictional time-independent Hamiltonian constructed out of the position of the extrema of amplitude. 
    \end{itemize}
    \begin{itemize}
    \item Another motivation for our work is the ``$S$-matrix bootstrap" program \cite{kruczenski2022snowmasswhitepapersmatrix,correia2020analyticaltoolkitsmatrixbootstrap, Bhat:2023puy}, which has been extremely successful in constraining physical $S$-matrices. In this context, $S$-matrices are studied based on the set of global symmetries, and other physical properties. $S$-matrices have information of all possible interactions relating free asymptotic ``\textit{in}" and ``\textit{out}" states. Simplest representation of the $S$-matrix operator is $S = \Omega (\infty)^{\dagger} \Omega (-\infty)$, where $\Omega(t)$ relates the ``in" and ``out" states to the eigenstates of the free Hamiltonian in $t=\mp \infty$ \footnote{If the total Hamiltonian is $H=H_0+V$ (``in" and ``out" states $\Psi_{\alpha}^{\pm}$ are eigenstates of this Hamiltonian) with $H_0$ being the free Hamiltonian (eigenstates $\Phi_{\alpha}$). They are related through $\Omega(t)==e^{i H t}e^{-i H_0 t}$ as $\Psi_{\alpha}^{\pm}=\Omega(\mp) \Phi_{\alpha}$. This relation holds only if the ``in" and ``out" states are smooth superposition of different $\Psi_{\alpha}$s. Note that this superposition is similar to the notion of TFD state.}. We expect that the fictional Hamiltonian we work with is sensitive to all possible asymptotic states of a theory and this in principle can illuminate the relationship between properties such as unitarity and analyticity of amplitudes, to the properties of complexity.
\end{itemize} 
In \cite{Bianchi:2022mhs}, it was shown that given a scattering amplitude $\mathcal{A}(z)$, where $z$ is an input (for example, the scattering angle) which we can vary, the position of the zeroes $(z_i)$ of the logarithmic derivative with respect to the position coordinate $z$,
\begin{equation}\label{derivative}
   \mathcal{F}=\frac{d Log[\mathcal{A}]}{d z}=0,
\end{equation}
can be treated as the eigenvalues of a fictional Hamiltonian. With this assumption, one can then study the level spacing ratios of this fictional Hamiltonian \cite{Bianchi:2022mhs, Bianchi:2023uby},
 \begin{equation}\label{LS}
     r_n=\frac{z_{n+1}-z_n}{z_n-z_{n-1}}=\frac{\delta_{n+1}}{\delta_n}, \mathcal{R}_n=min\{r_n, \frac{1}{r_n}\},
\end{equation}
 
 level statistics, and the scattering form factor \cite{Bianchi:2024fsi} 
 \begin{equation}\label{ScFF}
 ScFF= \frac{1}{L^2}\langle\sum_{z_i,z_j}e^{i s (z_i-z_j)}\rangle.
 \end{equation}
\begin{figure}
 \centering
    \includegraphics[width =  \linewidth]{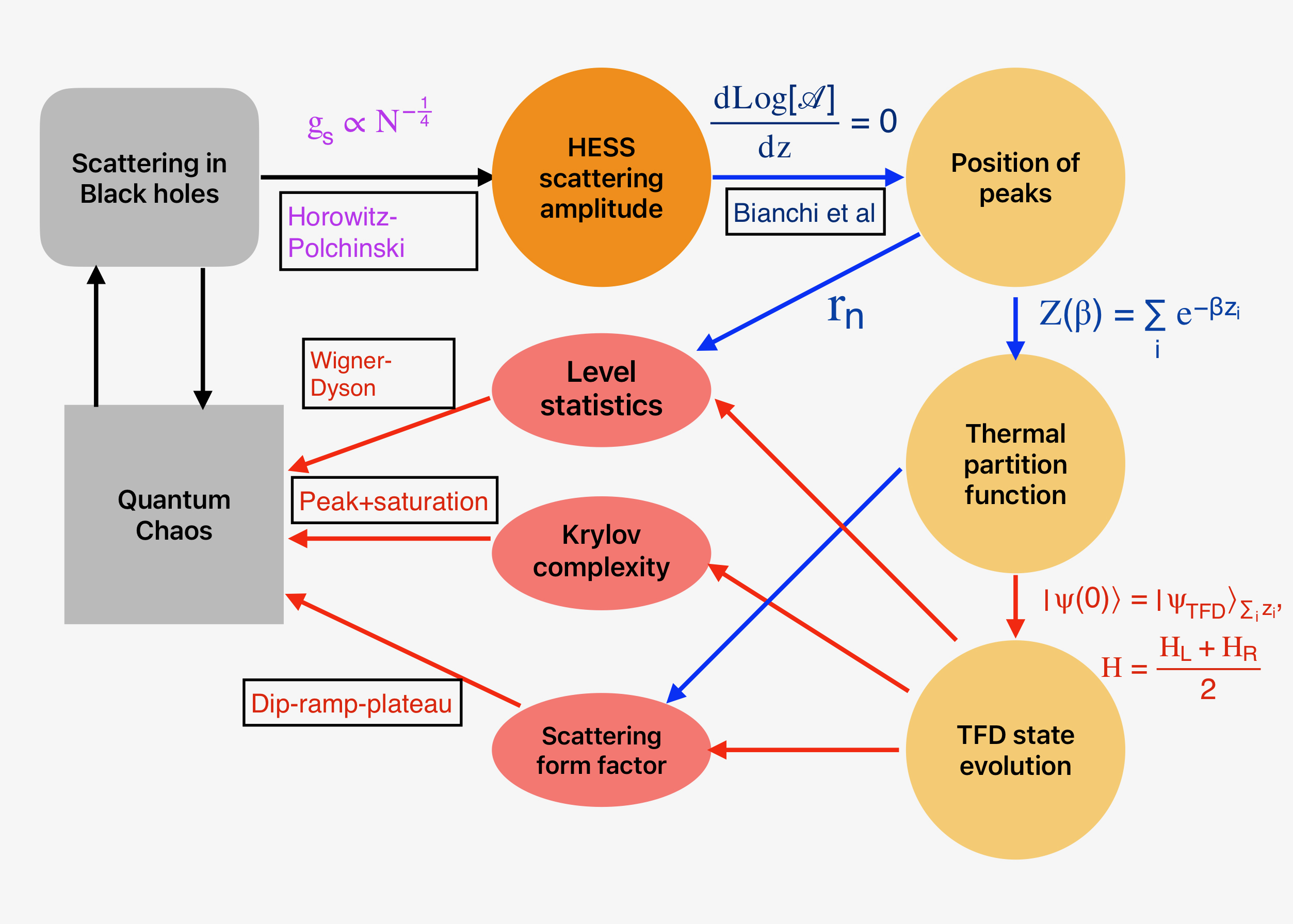}
    \caption{Graphical representationof our results. The \textcolor{blue}{blue} arrows represent previous studies of chaotic  scattering amplitudes through level statistics and scattering form factor. The \textcolor{red}{red} arrows correspond to the new directions we have explored in this letter. Our method can give rise to all three probes of chaotic scattering, namely, level statistics, form factor and complexity. All of these probes, when studied for the HESS scattering, indicate that the black hole scattering is quantum chaotic.}
    \label{fig:graphic}
\end{figure}
 
 If we truncate the process to finite number of such zeroes ($L$) and study the distribution statistics, it was shown in \cite{Bianchi:2024fsi} that these quantities capture the information of chaos in scattering amplitudes. While the level statistics for chaotic amplitudes follow Wigner-Dyson distribution, the scattering form factor show the dip-ramp-plateau behavior typical to the spectral form factor of Gaussian ensembles from the random matrix universality class. Since the zeroes of $\mathcal{F}$ in Eq. \eqref{derivative} are treated as eigenvalues of a Hamiltonian, it is easy to construct a thermal partition function from them as $Z(\beta)= \sum_{i} e^{-\beta z_i}$. This notion of partition function built from scattering amplitude motivates us to map the scattering problem to a thermo-field double (TFD) state evolution under the Hamiltonian devised such that it shares the same eigenvalues as the position of the extrema of a given scattering amplitude. We consider a general TFD initial state 
 
 \begin{equation}
 |\psi(s=0)\rangle= \frac{1}{\sqrt{Z(\beta)}}\sum_i e^{-\beta E_i/2} |i\rangle_L|i\rangle_R,
 \end{equation}
 and its evolution through the fictional Hamiltonian 

 \begin{equation}
 H_f=\frac{1}{2}(H_L\otimes I_R+I_L\otimes H_R),
 \end{equation}
 where $\beta$ is the inverse temperature in the thermal description, and the eigenvalues of $H_L, \, H_R$ are $E_i=z_i$ (solutions of Eq. \eqref{derivative}). This ensures a one-to-one map between the position of the extrema of an amplitude and a TFD state evolution under the explicitly built-out $H_f$ \footnote{It is worth noting here that one could equivalently consider the initial state to be a superposition of the eigenstates of a Hamiltonian $H$ sharing same eigenvalues as the position of the zeros of $\mathcal{F}$. Let us assume that these eigenstates are $|i\rangle$ and the initial state is $|\psi(s=0)\rangle=\frac{1}{\sqrt{L}}\sum_{i=1}^{L} |i\rangle$ and evolve it with respect to the Hamiltonian written as a $\left(L\times L\right)$ diagonal matrix with diagonal entries being the position of the zeros $\left(H=Diag\left(z_i\right)_L\right)$. This would give rise to a similar complexity profile if we take the definition of inner product is defined thermally with the $e^{-\beta H}$. The reason for this is that the nontrivial TFD evolution we study can be equivalently represented by one of two copies remaining a spectator (let's say the left one) and only the other one changes non-trivially with respect to $H_R$. However, dealing with TFD states generalizes the idea to thermal states in one of the two copies as well. Also we chose this way of describing due to computational reasons and having a more natural reason of including the thermal factor $e^{-\beta H}$ than just keeping it as a choice of inner product. This natural introduction of $\beta$ enables one to consider complex angular momentum.}. This TFD state construction can be naively thought of as equivalent to the smooth superposition of the energy eigenstates of the full interacting Hamiltonian $H=H_0+V$ (with $V$ being the interaction part of the Hamiltonian), usually considered as ``in" and ``out" states in QFT description of scattering. Hence, the state evolution we consider should respect asymptotic symmetries of the scattering theory under scrutiny. 
 
 \noindent Although this fictional evolution of the TFD state can equivalently represent a given amplitude, the coordinate $s$ in Eq. \eqref{ScFF} is not the actual time coordinate. Rather, it can be interpreted as the angular momentum or partial wave number, which becomes a continuous variable in the large spin limit in cases where $z$ is scattering angle (see section $4.3$ of \cite{Bianchi:2024fsi}). Also, in the description of the TFD, the usual inverse temperature $\beta$ should be identified as the complex part of the angular momentum $s$ \footnote{This is exactly similar to how complex time is identified with the inverse temperature $\beta$ in thermal setups.}, if it has any. Since this formalism does not have any time-dependence, and the evolution is itself modelled with respect to the angular momentum instead of time, we expect it to capture the information of scattering for $t \rightarrow \pm\infty$ and provide us insights on asymptotic states of the theory as well as the kind of interaction. Now, let us discuss how to compute the complexity of such an evolution given the thermal partition function.

\paragraph*{\textbf{General strategy}.---}
 To study the complexity in the Krylov basis, one needs to firstly compute the Lanczos coefficients $a_n$, and, $b_n$, which provide a tridiagonal representation of the Hamiltonian and depends on the choice of the initial state \cite{Balasubramanian:2022dnj, Nandy:2024htc}. The problem then becomes that of a one-dimensional Markov chain where one can compute the wave-functions $\phi_n(s)$ at a parameter value $s$ in each Krylov basis vector $|K_n\rangle$. These wave-functions can be derived by solving a recursive differential equation
 \begin{equation}
     i \frac{d\phi_n(s)}{ds}= a_n \phi_n(s)+b_n \phi_{n-1}(s)+b_{n+1}\phi_{n+1}(s),
 \end{equation}
 with the boundary condition $\phi_n(s=0)=\delta_{n,0}$ and knowledge of the Lanczos coefficients. Finally, the Krylov complexity is defined as the average position of the state in the Krylov basis.

 \begin{equation}
     C(s)=\sum_n n |\phi_n(s)|^2.
 \end{equation}
 Hence, given an initial state and Hamiltonian, the crucial information needed to compute the complexity are the Lanczos coefficients. There are two ways to derive these coefficients. Firstly, starting from an initial state, which is the first vector $|K_0\rangle$ in the Krylov basis, it is possible to derive the Lanczos coefficients by explicitly constructing the Krylov basis vectors $|K_n\rangle$ using the Lanczos algorithm (see Supp. Mat. \ref{lanc}). Alternately, one can start from the survival amplitude $S(s)=\langle \psi(s)|\psi(0)\rangle$ and derive its moments ($\mu_j$) by taking $j$-th derivative with respect to $s$ at $s$ going to zero \cite{Parker:2018yvk, Balasubramanian:2022tpr},
 \begin{equation}
     \mu_j=\frac{d^j}{ds^j} S(s)|_{s=0}=\langle K_0|(iH_f)^j|K_0\rangle, \, \text{for}\, j\in \boldsymbol{Z}^+,
 \end{equation}
and solve a moment recursion relation to get the Lanczos coefficients. 

Now, for the TFD state evolution we consider, the survival amplitude can be written simply as 
\begin{equation}
    S(s)=\frac{Z(\beta-i s)}{Z(\beta)}
\end{equation}

Hence, given the numerically derived partition function for each scattering amplitude, we can formally derive the Lanczos coefficients using the so-called moment recursion relation, and compute complexity. This formalism of computing complexity is therefore general and valid to any general scattering amplitude as long as it has resonances. 

\begin{figure}
    \centering
    \includegraphics[width = 0.49\linewidth]{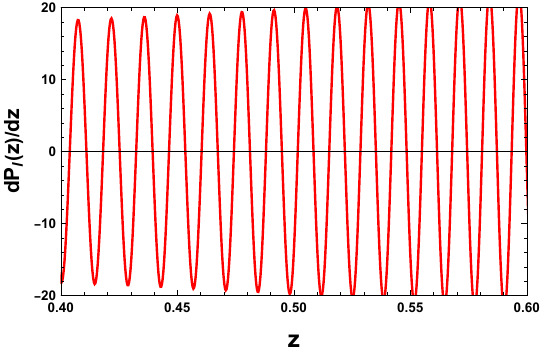}
    \includegraphics[width = 0.49\linewidth]{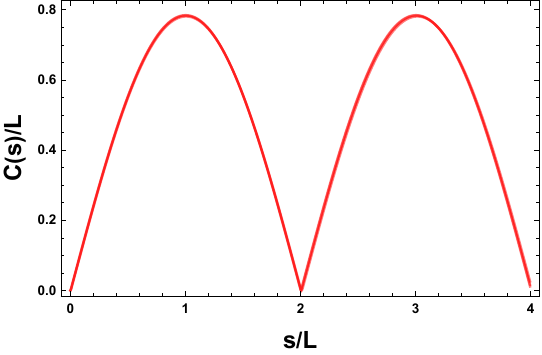}
    \includegraphics[width = 0.49\linewidth]{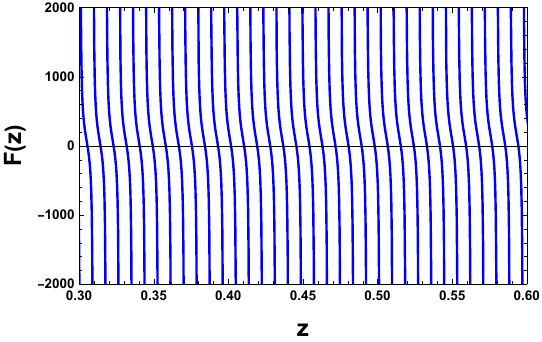}
    \includegraphics[width = 0.49\linewidth]{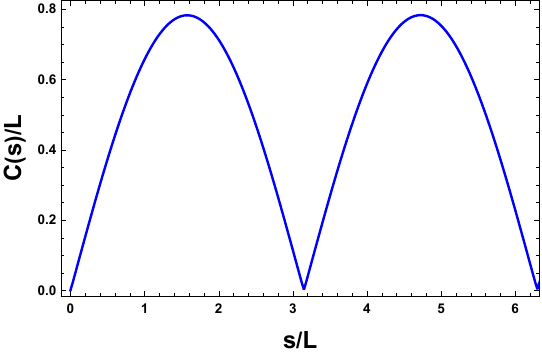}
\caption{\textbf{Non-chaotic:} Upper-Left: Plot showing the location of zeros of the Legendre polynomial (with \(\ell=400\)). Upper-Right: Complexity for leading Regge trajectory. Lower-Left: Plots showing the zeros in logarithmic derivative of the Veneziano amplitude (with \(s_1=225.1/\alpha'\)). Lower-Right: Complexity for the Veneziano amplitude. \(L\) denotes the number of eigenvalues in each case. For the decay of the leading Regge trajectory \(L=N=400\) and for the Veneziano amplitude \(L=228\) with \(s_1=225.1/\alpha'\).}
    \label{fig:Non-chaotic}
\end{figure}

\paragraph*{\textbf{Connections to black-hole scattering.}--}Using this general methodology, we compute complexity of the following scattering amplitudes. We have two examples each for non-chaotic and chaotic cases. As non-chaotic examples, we report the complexity for i) the leading Regge trajectory of highly excited string state(HESS) scattering into two tachyons, and, ii) the first ever known string scattering amplitude, namely, the 4 Tachyon scattering Veneziano amplitude. On the other hand for the chaotic examples, we study the complexity of a generic HESS scattering into i) two and ii) three Tachyons respectively, both of which are examples of chaotic scattering amplitudes as pointed out in \cite{Gross:2021gsj}. These amplitudes were first studied in using the Del Giudice-Di Vecchia-Fubini (DDF) formalism \cite{DELGIUDICE1972378}. It is also worth noting that the chaotic examples are classic cases of the string-black hole correspondence by Susskind-Horowitz-Polchinski \cite{Susskind:1993ws, Horowitz:1996nw, Polchinski:2015cea}, where strings at high energies (small string coupling $g_s \propto N^{-1/4}$, with high excitation level $N$, where the string length $\ell_s$ is of the same order as its Schwarzchild radius $2 G M$, with $M \propto \sqrt{\frac{N}{\alpha'}}$) are dual to a black hole. Therefore, the chaotic behavior observed in these examples supports the presence of chaotic scattering in black holes. While the semiclassical realization of chaos in black holes is derived from the study of the Lyapunov exponent using out-of-time-ordered correlators \cite{Maldacena_2016}, our notion of chaos, based on the complexity of scattering amplitudes, provides a quantum field theoretic realization of chaos of black hole $S$-matrices in case of a string-black hole transition. The main takeaway of this letter is graphically represented in Fig. \ref{fig:graphic}. All the numerical results we report in the main text are for $\beta=0$, which corresponds to infinite temperature TFD from state evolution perspective and purely real angular momentum $s$ from scattering perspective, (the dependence on $\beta$ can be found in Fig. \ref{fig:RieZetaplots} of supplemental material \ref{examp}).

\paragraph*{\textbf{Non-chaotic examples.}---} The simplest way to understand what we mean by non-chaotic amplitudes is again through the position of its resonances with respect to the scattering angle. If the resonances of the amplitudes, and hence, the position of the zeroes of Eq. \eqref{derivative} are equispaced, then it becomes completely predictable and we expect that to show up in the behavior of complexity as well. We study two such examples below.

\subsection*{Leading Regge trajectory of HESS to two Tachyons (HTT)}

The decay amplitude of a highly excited string state at the leading Regge trajectory (that is, a string state with mass level \(N\) equal to its spin \(\ell\)) and with helicity \(J=0\) to two tachyons is proportional to \(P_{\ell}(\cos\theta)\) (where \(\theta\) is the angle of scattering in the center of mass frame, the angle of outgoing particles with the spin of the decaying particle)\footnote{For further details, please see section $4.3$ of \cite{Bianchi:2024fsi}.}. For \(\ell=N>>1\), we have \cite{gradshteyn2007},
\begin{equation}
    \frac{dP_\ell (\cos\theta)}{d\theta} \propto \sin\left((\ell+\frac{1}{2})\theta - \frac{\pi}{4}\right),\,\,\text{ for } 0<\theta<\pi
\end{equation}
Therefore, the zeros are located at,
\begin{equation}
    \theta_k = \pi \frac{k+\frac{1}{4}}{\ell+\frac{1}{2}},\,\,\text{ for } k = 1,\dots,\ell
\end{equation}
Here it is explicit that the zeroes are equally spaced (See top left of Fig. \ref{fig:Non-chaotic}). We show the complexity of this particular scattering amplitude computed according to our proposal in top right of Fig. \ref{fig:Non-chaotic}. We find a periodically oscillating behavior of the complexity.

\subsection*{Veneziano Amplitude}

The Veneziano amplitude is the first ever studied string amplitude which describes the 4pt scattering of 4 tachyons in open Bosonic string theory \cite{Veneziano:390478}. This amplitude is simply the Euler Beta function,
\begin{equation}
    \mathcal B(-\alpha' s_1 - 1, -\alpha' s_2 - 1) = \frac{\Gamma(-\alpha' s_1 - 1)\Gamma(-\alpha' s_2 - 1)}{\Gamma(-\alpha' s_1 -\alpha' s_2 - 2)}
\end{equation}
(here $\alpha'$ is related to the inverse string tension and the string length is given by $\ell_s = \sqrt{2\alpha'}$).

\noindent The scattering angle \(\theta\) in the center of mass frame can be obtained by \(z=\cos\theta = 1 + 2s_2/(s_1+4/\alpha')\) (the mass squared of the tachyon is given by \(-1/\alpha'\)). We look for zeroes of the logarithmic derivative of the above amplitude at fixed center of mass energy \(s_1\) as a function of the variable \(z\) and treat them as our eigenvalues (bottom left of Fig. \ref{fig:Non-chaotic}). Then we proceed to calculate the scattering complexity. We observe that the complexity for the Veneziano amplitude is again a periodically oscillating function (bottom right of Fig. \ref{fig:Non-chaotic}). This periodic nature, of course, arises due to the equal spacing of the zeros. This property holds true even in the high energy limit of the Veneziano amplitude, and thus insensitive to low energy modifications.

\paragraph*{\textbf{Chaotic examples.}----}
Chaos in scattering amplitudes is marked by erratic behavior with respect to the scattering angle. HESS scattering into two or three tachyons exemplifies this. It is also worth noting that these two chaotic examples can be written as just the non-chaotic examples studied in the last section multiplied by certain dressing  factors \cite{Gross:2021gsj}. We analyze the positions of resonances, treating them as Hamiltonian eigenvalues, and study the TFD state evolution under this Hamiltonian. This approach aims to determine if the complexity profile reveals characteristics of chaos. \\
\noindent In the DDF formalism a generic highly excited string state at level \(N\) (and mass given by, \(\alpha'M^2 = N - 1\)) is constructed by scattering photons successively from a tachyon momenta \(\Tilde{p}\) (with \(\Tilde{p}^2 = +1/\alpha'\)). In this formalism the string coherent vertex operator can be constructed relatively easily for highly excited string state (\(N>>1\)) where it can be very tedious to construct the vertex operators by brute force solution of the Virasoro constraints. All the photons that are to be scattered are chosen to have momentum parallel to a reference null momentum \(q\). Suppose \(g_n\) number of photons are scattered with momenta \(n q\) for \(n=1,2,\dots\), then the final state will have momentum \(p = \Tilde{p} - N q\) where \(N = \sum_n n g_n\) and helicity \(J = \sum_n g_n \). To satisfy the mass condition \(p^2 = - (N-1)/\alpha'\) one chooses \(\Tilde{p}\cdot q = 1/2\alpha'\). In fact, any excited string state can be constructed in this way \cite{Gross:2021gsj}. Scattering amplitudes involving these highly excited string states in Bosonic string theory are computed in \cite{Gross:2021gsj,Bianchi:2023uby,Das:2023cdn} .

\begin{figure}[htbp]
    \centering
    \includegraphics[width = 0.49\linewidth]{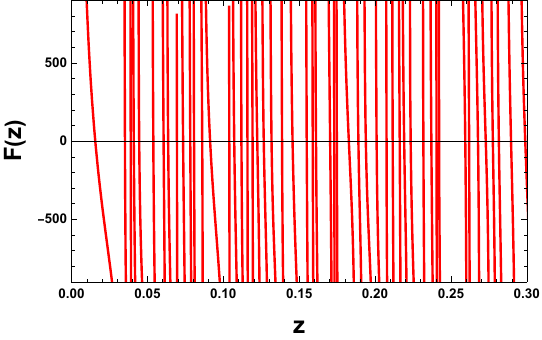}
    \includegraphics[width = 0.49\linewidth]{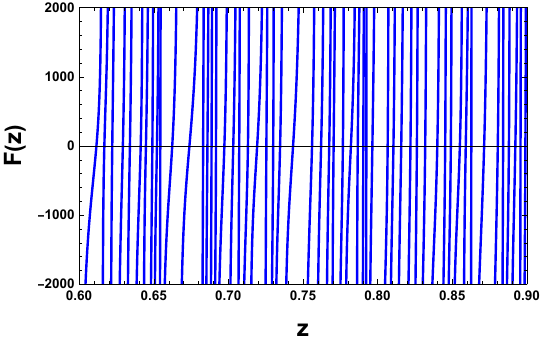}    
    \vskip\baselineskip    \includegraphics[width = 0.85\linewidth]{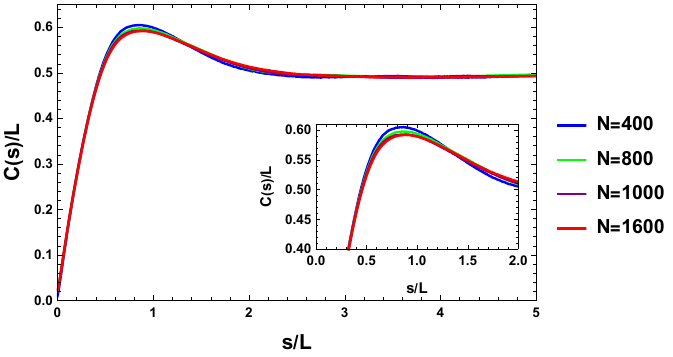}
    
\vskip\baselineskip
\includegraphics[width = 0.85\linewidth]{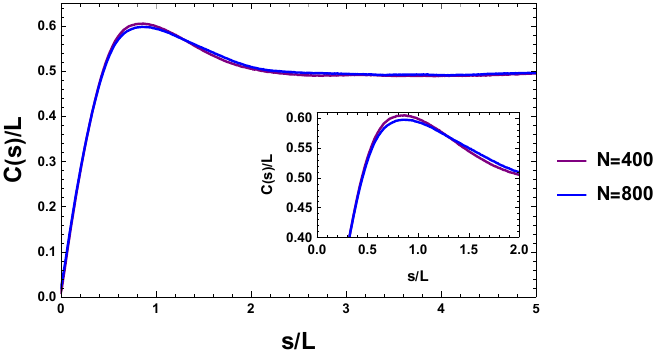}
\caption{ \textbf{Chaotic:} Top left: Zeroes of $\mathcal{F}$ for HESS decaying to 2  Tachyons,Top right: Zeroes of $\mathcal{F}$ for HESS decaying to 3  Tachyons. Middle: Complexity for the 3pt scattering of HESS and 2 Tachyons. Bottom: Complexity for the 4pt scattering of HESS and 3 Tachyons. For the computation of complexity the `eigenvalues' in each case is re-scaled and shifted to fall in the range \(-2\) to \(2\) to indicate the somewhat universal behavior of the complexity profile. As before, \(L\) denotes the number of eigenvalues in each set. Note the peak and saturation at~$0.5$ of the complexity in chaotic cases.}
    \label{fig:Chaotic}
\end{figure}

\subsection*{HESS to two Tachyons (HTT)}

The amplitude for the decay of a HESS at mass level \(N\), helicity \(J\) and a particular partition \(\{g_n\}\) is proportional to,
\begin{equation}
\displaystyle
    (\sin\alpha)^J \prod_{n=1}^N \left(\sin\left(\pi n \cos^2 \frac{\alpha}{2}\right) \frac{\Gamma(n\cos^2\frac{\alpha}{2})\Gamma(n\sin^2 \frac{\alpha}{2})}{\Gamma(n)}\right)^{g_n}
\end{equation}

(here $\alpha$ is the angle between the outgoing tachyons and the DDF photon momentum $q$).\footnote{For more details, please have a look at \cite{Gross:2021gsj}.}

We can find the zeros of logarithmic derivative of the above expression as a function of $z=\cos^2\frac{\alpha}{2}$ (plot shown in top left of Fig. \ref{fig:Chaotic}). Calculating complexity for different choices of $N$, we observe a clear peak before saturation for all the cases, which according to our proposal is a signature of chaotic scattering (refer to Fig. \ref{fig:Chaotic}). While these plots are average over all possible partitions $\{g_n\}$ for each $N$ \footnote{For producing the plots, an averaging procedure is done by choosing different \(J\) randomly for a given \(N\), see \cite{Bianchi:2023uby} for a detailed discussion on the probability distribution of the highly excited string states constructed from DDF formalism.}, it is interesting to note that each partition individually gives rise to the same behavior of complexity (see Fig. \ref{fig:singleRealizationplots} of Supp. Mat. \ref{LS&SFF}).

\subsection*{HESS to three Tachyons (HTTT)}
Now we proceed to probe chaos in the scattering amplitude of one HESS and 3 tachyons (which was found to be chaotic from level-statistics analysis) from our scattering complexity perspective. For simplicity we consider the case where all the scattered photons have equal circular polarization and take the amplitude in the Regge limit $s_1 >> |s_2|$ (the amplitude in the fixed angle high energy scattering can also be analysed in a similar manner \cite{Bianchi:2023uby}),
\begin{equation}
\begin{aligned}
    &\Gamma\left(-\frac{s_2}{2}-1\right) s_1^{\frac{s_2}{2}+1} \left(-\sqrt{s_1}\left(1-\frac{1}{2}\sin\theta\right)\right)^J\\
    &\times \prod_{n=1}^N \left(\frac{\Gamma(n\sin\theta) \Gamma(n-n\sin\theta)}{\Gamma(n)}\sin(n\pi\sin\theta)\right)^{g_n},
\end{aligned}
\end{equation}
(here $\theta$ is the angle between the incoming momenta and the outgoing momenta in the center of mass frame)

We find the zeros of the logarithmic derivative of the dressing factor which comes with the Regge limit of the Veneziano amplitude, with respect to the variable $z=1/(2\cos^2(\theta'/2))$ where $\theta' = \pi/2 - \theta$ (top right of Fig. \ref{fig:Chaotic}). Computing scattering complexity we again find a saturation with a visible peak (bottom of Fig. \ref{fig:Chaotic}) which reinforces that scattering amplitudes involving generic HESS are always chaotic which can be linked to the fact that these HESS comprise black holes in string theory which are known to be chaotic semi-classically.

\paragraph*{\textbf{Main result.}---} 
This letter demonstrates that each scattering amplitude can be mapped to a uniquely devised Hamiltonian evolution of a thermofield double (TFD) state, enabling the computation of complexity in the Krylov basis. This mapping provides a framework for a complexity-based classification of scattering amplitudes and connects the concepts of chaos in quantum state evolution \cite{Parker:2018yvk, Balasubramanian:2022tpr}, and scattering amplitudes \cite{Rosenhaus:2020tmv, Gross:2021gsj, Bianchi:2024fsi}. Chaotic scattering amplitudes display a complexity profile characterized by early growth, a peak, and smooth saturation, mirroring chaotic Hamiltonian evolution. In contrast, non-chaotic amplitudes exhibit periodic behavior without a peak. Since the whole formalism is independent of time, this notion of complexity is ideal to be associated to the asymptotic states of a $S$-matrix from $t_i=-\infty$, and, till $t_f=+\infty$, and respect the asymptotic symmetries.

Our study reveals that chaotic scattering complexities show a characteristic peak even for individual realizations, unlike the level statistics and scattering form factors (see Supp. Mat. \ref{LS&SFF} for more details) where averaging is required for meaningful insights on chaos \cite{Bianchi:2024fsi}. This indicates that complexity is a more robust probe of chaos as compared to level statistics or scattering form factor. These findings further link the properties of chaotic scattering amplitudes to Krylov complexity, with high-energy string state scattering indicating chaos in black hole scattering. This signifies that slight changes made to the input state going into a black hole can drastically alter the relative complexity of the output state, highlighting the randomness of information scrambling within black holes. 

\paragraph*{\textbf{Future directions.}---}Future directions for this work include i) applying our methods to general scattering amplitudes, such as using a truncated partial wave expansion and modifying low-energy data to test the sensitivity of complexity under such changes. ii) We would also like to build a precise map between the properties of complexity and various properties of scattering amplitudes, namely, unitarity, analyticity, crossing symmetry etc in future. It will be further exciting to consider the possibility of constraining the set of allowed $S$-matrices from complexity along the lines of \cite{Bose:2020cod, Bose:2020shm}. Additionally, iii) exploring the question of thermalization in scattering amplitudes by studying multiple high-energy string states scattering \cite{Das:2023cdn, Rabinovici:2021qqt} could enhance our understanding of HES states. iv) Investigating the relations between thermalization and complex momentum $\beta$ from the TFD description, as well as identifying scattering processes that produce complexity profiles with saturation without peak (characteristic of integrable systems) remain as intriguing avenues for further research in this direction. Finally, v) it would be great to check if these Lanczos coefficients derived from amplitudes can reproduce the Greens function \cite{Parker:2018yvk}, and study diffusion coefficients for various scattering processes. This also raises the question whether studying resonances in black hole quasinormal modes can also provide with similar insights about complexity and chaos.

\begin{acknowledgments}
\vspace{3em}
\noindent The authors would like to thank Prashanth Raman, Aninda Sinha, and, Jacob Sonnenschein for useful discussions and comments on the draft. We would also like to thank Parthiv Haldar for collaboration in the initial stages of the project and ongoing collaboration. The work of A.B.~is supported by the Polish National Science Centre (NCN) grant 2021/42/E/ST2/00234. A.B.~would like to thank International Centre for Theoretical Sciences, Bengaluru, and, the organizers of the program ``Quantum Information, Quantum Field Theory, and Gravity" for the hospitality when this work was in the final stage.
\end{acknowledgments}

\newpage
\onecolumngrid
\renewcommand{\theequation}{S.\arabic{equation}}
\setcounter{equation}{0}

\section*{Supplemental Material}
\section{Lanczos algorithm }\label{lanc}
Let's say we start from an initial state $|\psi(0)\rangle$. The Schrodinger evolution of this state in time under a Hamiltonian $H$ is
\begin{equation}
    |\psi(t)\rangle=e^{-iHt} |\psi(0)\rangle=|\psi(0)\rangle-i t H|\psi(0)\rangle+\frac{i^2t^2}{2}H^2|\psi(0)\rangle+\cdots.
\end{equation}

In perturbative series expansion in time, one can notice that higher order terms in this series comes from higher powers of the Hamiltonian, $H^n$, on the initial state.
However the states associated to each power of $t$ in such a series expansion, are not orthonormal to each other. Therefore, to define measures consistently, one needs to orthonormalize these vectors in the time-series expansion with respect to each other. This is done by the Lanczos algorithm. The resulting orthonormal basis is known as the Krylov basis. 

Lanczos algorithm, apart from generating a set of orthonormal basis vectors, also provides a tri-diagonal form of the Hamiltonian. Historically, the tridiagonal matrix derived by using the algorithm is well known to approximate largest eigenvalues of the original Hamiltonian.

The basic algorithm is as follows. 
\begin{itemize}
    \item Identify the initial state as the first vector $|K_0\rangle$ of the Krylov basis. $|\psi(0)\rangle=|K_0\rangle$, and generate next basis vectors by the action of the Hamiltonian.  $|A_1\rangle=H|K_0\rangle- a_0|K_0\rangle,\, a_0=\langle K_0|H|K_0\rangle.$

    \item $b_0=0,\, b_1=\sqrt{\langle A_1|A_1\rangle}.$

    \item $|A_{n+1}\rangle=H|K_n\rangle-a_n|K_n\rangle-b_n|K_{n-1}\rangle,\,a_n=\langle K_n|H|K_n\rangle.$

    \item $b_{n+1}=\sqrt{\langle A_{n+1}|A_{n+1}\rangle}$.

    \item Stop the recursive process once $b_m=0$, for some $m$. This happens when the full state Hilbert space is explored in the Krylov basis.
\end{itemize}

\textbf{Alternate way to compute Lanczos coefficients:} The alternate way to compute the Lanczos coefficients is to start from the so-called Survival amplitude, which is given by the overlap between the final and the initial states. 
\begin{equation}
    S(t)=\langle \psi(t)|\psi(0)\rangle.
\end{equation}
The moments of the survival amplitude can be then related to the possible amplitudes in one dimensional Markov chain, where $n$-th moment provides the expectation value of the $n$-th power of the Hamiltonian $H^n$ with respect to the initial state $|\psi_0\rangle=|K_0\rangle$. In the Markov chain, this expectation value $\langle K_0|H^n|K_0\rangle$ corresponds to the amplitude of returning to initial ($1$st) level by considering all possible paths/processes upto $n$-th level where points in the same level are connected through the $a_n$ amplitudes, and nearest neighbours of $m$-th and $(m+1)$-th level are connected through the diagonal path with coefficient $b_m$ (see Fig. $1$ in \cite{Balasubramanian:2022tpr}). Finally this gives rise to a moment recursion relation in terms of the Lanczos coefficients $a_n$ and $b_n$ (see section V.A in the review \cite{Nandy:2024htc}).

After solving the moment recursion relation starting from the survival amplitude, one can therefore compute all the Lanczos coefficients. For us, the survival amplitude is given by that of the TFD state, which is just the ratio of the partition function for time $t$ and time $0$ respectively.

Once we have all the Krylov basis vectors $\{K_n\}$ or the set of Lanczos coefficients $a_n$ and $b_n$, we can write down the final state as $|\psi(t)\rangle=\sum_n \phi_n(t) |K_n\rangle$, with $\phi_n$'s denoting the wavefunction in the Krylov basis. They follow the following differential equation in terms of the Lanczos coefficients $a_n$, and $b_n$.
\begin{equation}
    i \frac{d \phi_n(t)}{dt}= a_n \phi_n(t)+b_n \phi_{n-1}(t)+b_{n+1}\phi_{n+1}(t),
\end{equation}
with the boundary condition given by $\phi_n(t=0)=\delta_{n,0}$.

For unitary evolution, they follow the identity $\sum_n |\phi_n(t)|^2=1$, and complexity is defined as the average position of the state in the Krylov basis, $C(t)=\sum_n n|\phi_n(t)|^2$.

\begin{figure}[htbp]
    \centering
    \includegraphics[width = 0.48\linewidth]{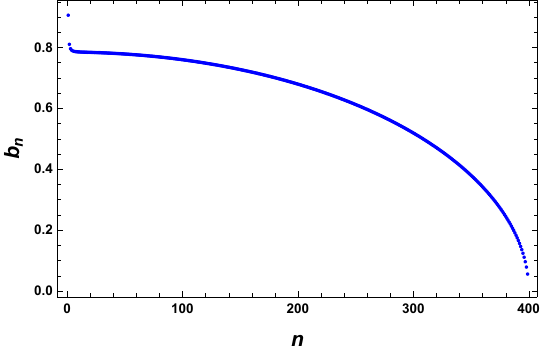}
    \includegraphics[width = 0.48\linewidth]{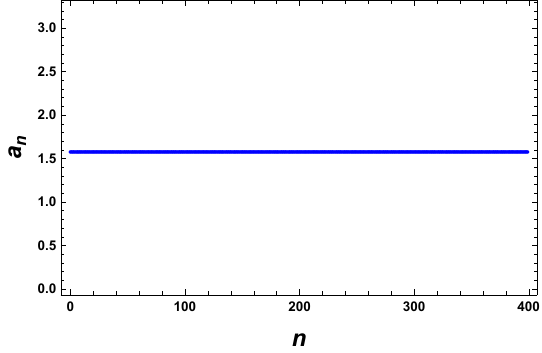}
    \includegraphics[width = 0.48\linewidth]{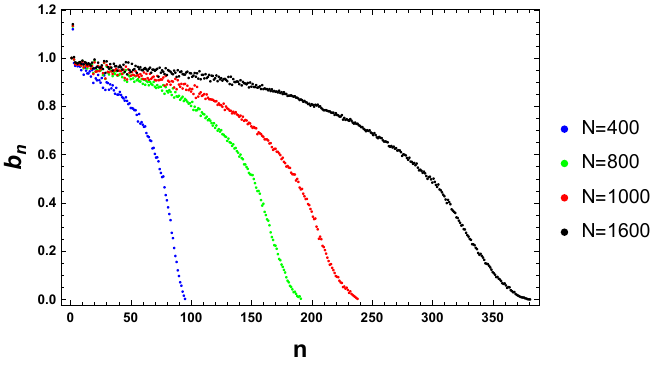}
    \includegraphics[width = 0.48\linewidth]{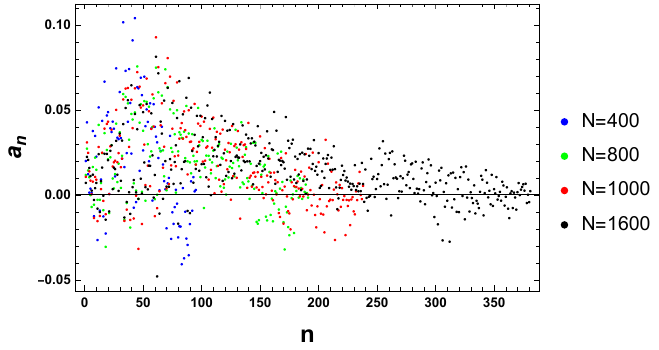}
    \caption{ \textbf{Lanczos coeff. for non-chaotic and chaotic cases:} Top left: $b_n$ vs $n$ for leading Regge trajectory of HESS to two Tachyons, Top right: $a_n$ vs $n$ for leading Regge trajectory of HESS to two Tachyons. Bottom left: $b_n$ vs $n$ for chaotic HESS to two Tachyons for increasing levels of string excitation $N$, Top right: $a_n$ vs $n$ for chaotic HESS to two Tachyons for increasing levels of string excitation. }
    \label{fig:anbnintChaotic}
\end{figure}

For state evolution, if one starts from a TFD state as the initial state, the distinction between chaotic and non-chaotic Hamiltonian evolution are listed below.

\begin{itemize}
    \item The profiles of the Lanczos coefficients in Schrodinger picture do not differ much when compared between chaotic and non-chaotic. However, a distinction can be made based on the variance, $Var(x)=\sqrt{\langle x^2\rangle-\langle x\rangle^2}$, of the Lanczos coefficients \cite{Balasubramanian:2024ghv}.

    \item In the complexity profile, the chaotic evolution always gives rise to a peak before saturation at a constant value. If we renormalize the time ($t_{ren}=\frac{t}{K}$) and the complexity ($C_{ren}=\frac{C}{K}$) by the Krylov dimension $(K)$, this saturation value is at $0.5$, which tells us that the maximum saturation value of Krylov complexity is $\frac{K}{2}$. However, the crucial distinction is provided by the peak above $0.5$ before the saturation, which is absent for non-chaotic evolution. There the behavior is either periodic and oscillatory, or a saturation without a peak. 
\end{itemize}

We have already discussed similar behavior of complexity found from the chaotic scattering amplitudes. In Fig. \ref{fig:anbnintChaotic}, we have shown the behavior of the Lanczos coefficients for one example of non-chaotic and chaotic example each. We notice that there is no clear qualitative difference between the Lanczos coefficients between the two cases. However, the complexity plots for the two cases differ substantially as reported in the main text (Figures \ref{fig:Non-chaotic} and \ref{fig:Chaotic}).

\section{Further examples}\label{examp}

\begin{figure}[htbp]
    \centering
   
    \includegraphics[width = 0.4\linewidth]{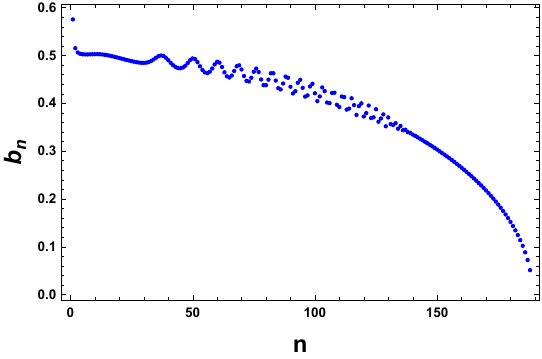}
    \includegraphics[width = 0.4\linewidth]{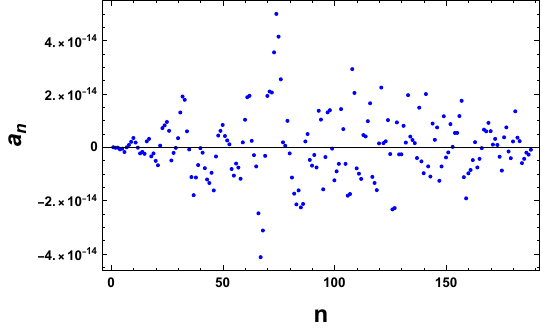}
    \vskip\baselineskip
     \includegraphics[width = 0.4\linewidth]{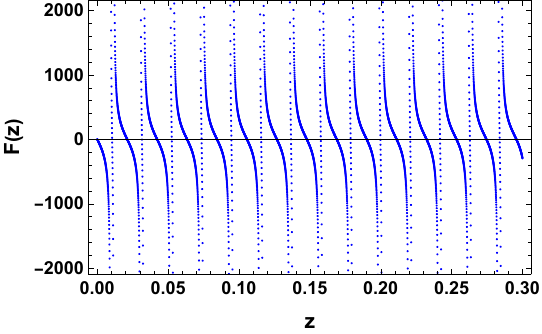}
    \includegraphics[width = 0.4\linewidth]{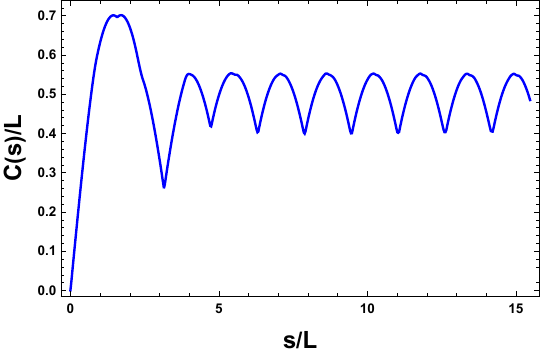}
    \caption{\textbf{Virasoro-Shapiro plots:} Scattering complexity for the Virasoro-Shapiro amplitude. Top-left: \(b_n \text{ vs } n\) plot. Top-right: \(a_n \text{ vs } n\) plot. Bottom-left: showing the equispaced zeros of the logarithmic derivative of Virasoro-Shapiro amplitude (at \(s_1=95.1\)). Bottom-left: the scattering complexity of Virasoro-Shapiro amplitude showing almost periodic behavior.}
    \label{fig:VirasoroShapiro}
\end{figure}

 \noindent \textbf{Virasoro-Shapiro (non-chaotic):} As a further example for non-chaotic scattering amplitude, we have got another well-known string amplitude, the Virasoro-Shapiro amplitude which describes scattering in closed string theory \cite{VS1, VS2, VS3}. In the following we consider Virasoro-Shapiro amplitudes for external massless states in terms of Mandelstam variables \(s_1,s_2\),
 \begin{equation}
     \frac{\Gamma(-s_1)\Gamma(-s_2)\Gamma(s_1+s_2)}{\Gamma(1+s_1)\Gamma(1+s_2)\Gamma(1-s_1-s_2)}
 \end{equation}
\noindent Then we calculate the logarithmic derivative of above function with respect to the cosine of scattering angle \(z=1+ 2 s_2/s_1\), at fixed \(s_1\). Using the zeros of the resultant function we compute complexity which shows a persistent periodic behavior. This oscillatory behavior, we believe, is a result of the evident pattern in the position of zeros.

\begin{figure}[htbp]
    \centering
    \includegraphics[width = 0.32\linewidth]{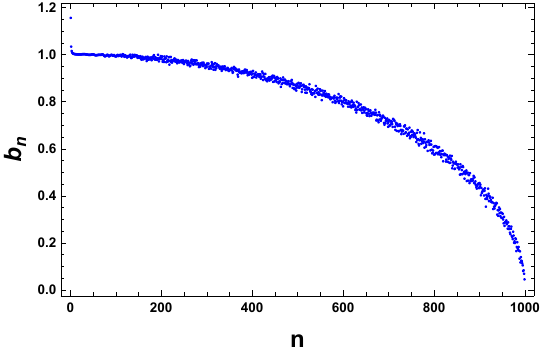}
    \includegraphics[width = 0.32\linewidth]{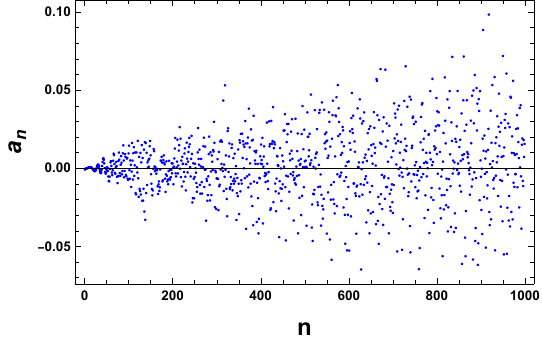}
    \includegraphics[width = 0.32\linewidth]{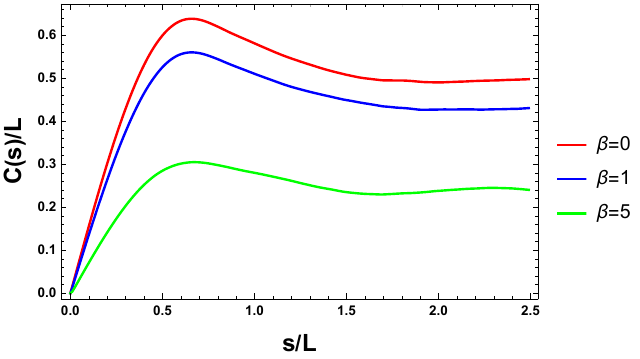}
    \caption{ \textbf{Leaky-torus and $\beta$-dependence:} Left: $b_n$ vs $n$ for imaginary part of non-trivial zeros of Riemann zeta function, Middle:$a_n$ vs $n$ plot for the same, Right: Complexity for the imaginary part of non-trivial zeros of Riemann zeta function and the dependence of complexity on the parameter $\beta$, which is put in by hand in the TFD state evolution formalism.}
    \label{fig:RieZetaplots}
\end{figure}

\noindent \textbf{Leaky torus (chaotic):} Here we report the behavior of complexity for the scattering in the leaky torus with poles at zeroes of the Riemann-Zeta function (well known to show random matrix universality)\cite{leakytorus}. 
This is one of the rare analytic example, unlike the string scattering amplitudes, where we know the position of the poles analytically. 

The leaky torus geometry is formed by cutting a piece of a hyperbolic space and identifying in the upper half plane a four points $y=\infty$ and $y=0; x=-1,0,1$ in sets of two (see \cite{Bianchi:2022mhs}, and \cite{Rosenhaus:2020tmv} for more details). Then one sends ingoing waves through the infinity of the cusp in the geometry to observe the phase shift of the outgoing waves. The phase shift $\Phi(k)$, with $k$ being the incoming wave momentum, in this case follows the following form

\begin{equation}
    \Phi(k)=\frac{Im[\zeta(1+2ik)]}{Re[\zeta(1+2ik)]}.
\end{equation}

The notion of chaos can again be understood when the phase shift is plotted against the incoming momentum (see Fig. 2 of \cite{Rosenhaus:2020tmv}). In \cite{Bianchi:2022mhs}, the authors studied the positions of $\Phi^{\prime}(k)=0$, and showed that it indeed follows a log-normal distribution close to Wigner-Dyson. More recently in \cite{Bianchi:2024fsi}, it was also shown that the scattering form factor after unfolding also shows the expected dip-ramp-plateau behavior typical to random-matrix-universality class GUE (Gaussian Unitary Ensemble). We therefore follow the same procedure as mentioned in our main text by using the position of the zeros of $\Phi^{\prime}(k)$ to constructa Hamiltonian that evolves a TFD state non-trivially. The behavior of the Lanczos coefficients and complexity computed from the Riemann-Zeta zeroes (Fig. \ref{fig:RieZetaplots}) indeed show the expected chaotic behavior with a peak before saturation at $0.5$.

 This by construction is an explicit realization of a Hamiltonian where the Riemann-Zeta zeroes are the eigenvalues. However, along the lines of \cite{Bhattacharya:2023yec}, one can argue that if we construct a tight binding Hamiltonian with these set of $a_n$, and $b_n$ coefficients as the hopping amplitudes, that would give rise to a physically meaningful example Hamiltonian with eigenvalues exactly equal to Riemann-Zeta zeroes. Hence, this tight-binding Hamiltonian can be the quantum mechanical Hamiltonian supporting the Hilbert-Polya conjecture. Furthermore, this example of leaky torus can be used to further the study of complexity for a class of scattering amplitudes described in \cite{Remmen:2021zmc}, and relate the properties of Riemann-Zeta function, for example, positive odd moments, meromorphicity, to that of complexity.

 \textbf{$\beta$-\text{ dependence}:} We further note the dependence of complexity in Fig. \ref{fig:RieZetaplots} on $\beta$, which corresponds to the complex angular momentum in case of scattering amplitudes. We find that with increasing $\beta$, the saturation value of the complexity comes down, and the peak becomes less pronounced. This is again qualitatively very similar to the inverse temperature dependence of complexity in state evolution as shown previously in \cite{Balasubramanian:2022tpr}. This further ensures a close correspondence between the inverse temperature in the state evolution and complex momentum in the scattering amplitudes. In case of the state evolution, the suppression of the complexity saturation can be understood by decreased contribution of higher energy eigenvalues for non-zero (and increasing) $\beta$ due to the factor $e^{-\beta E_n}$. In case of the scattering amplitudes, this means that extremas of scattering amplitudes positioned at higher values of scattering angle contribute less to the complexity if there is an imaginary part to the angular momentum. It would be interesting to explore this direction further to understand what this implies physically.

\section{Averaging over datasets: comparing level statistics, form factor and complexity}\label{LS&SFF}
The level statistics and scattering form factor for scattering amplitudes are defined in Eq. \eqref{LS} and \eqref{ScFF}. However, while studying them, and getting consistent plots, one needs to be a bit more careful. One needs to average over a significant number of data sets to get a good plot of scattering form factor (ScFF) which shows a clear dip-ramp-plateau behavior for chaotic data. A single realization typically fails to capture these characteristic features in a clear way even if there are \(\sim 1000\) data points in a data set. The same holds true for level-statistics also.  Here we exemplify this point by showing the ScFF and level-statistics of non-trivial zeros of Riemann zeta function without and with averaging over data sets.

\begin{figure}[htbp]
    \centering
    \includegraphics[width=0.42\linewidth]{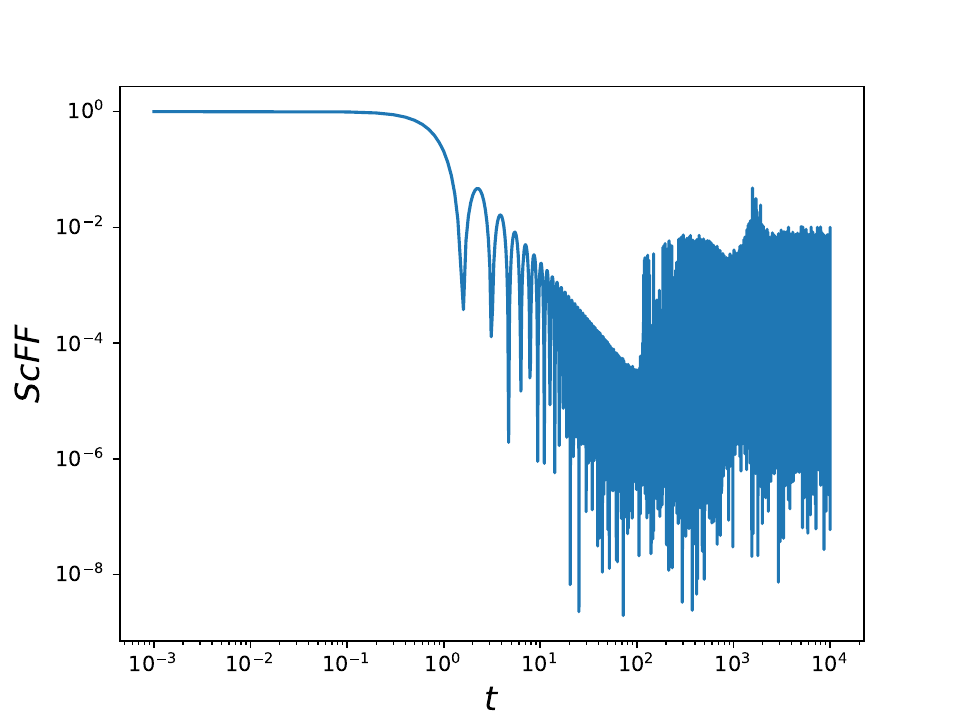}
    \includegraphics[width=0.42\linewidth]{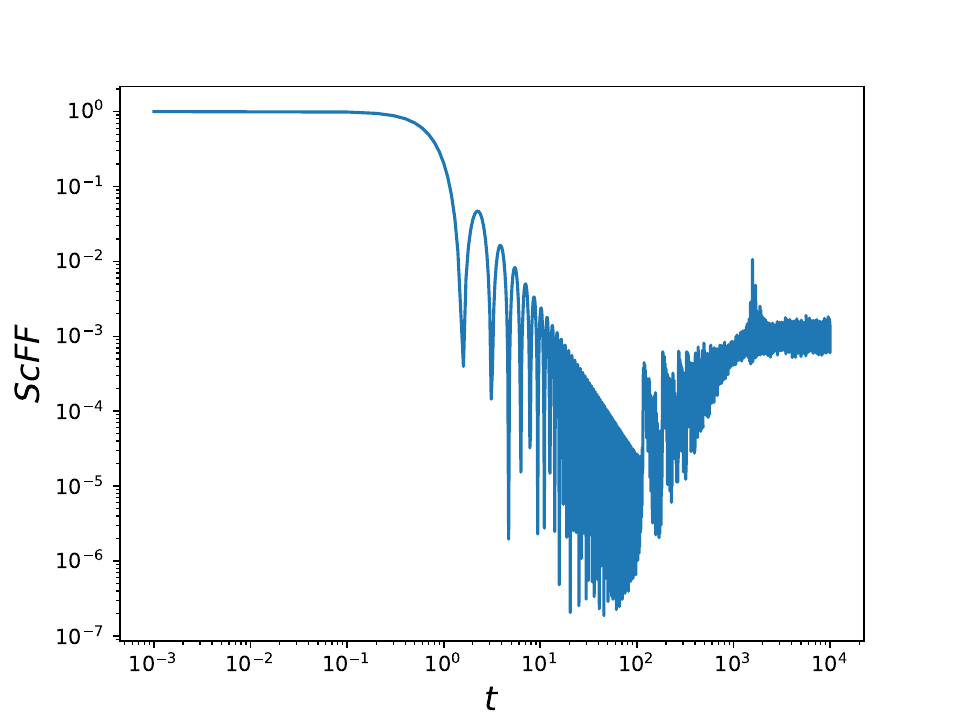}
    \vskip\baselineskip
    \includegraphics[width=0.42\linewidth]{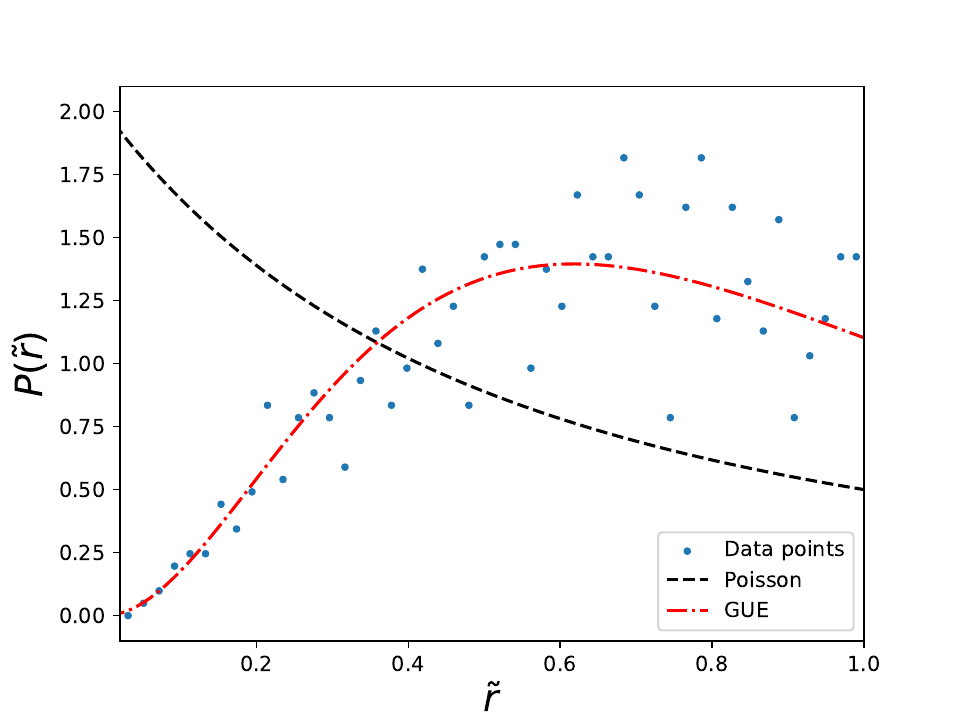}
    \includegraphics[width=0.42\linewidth]{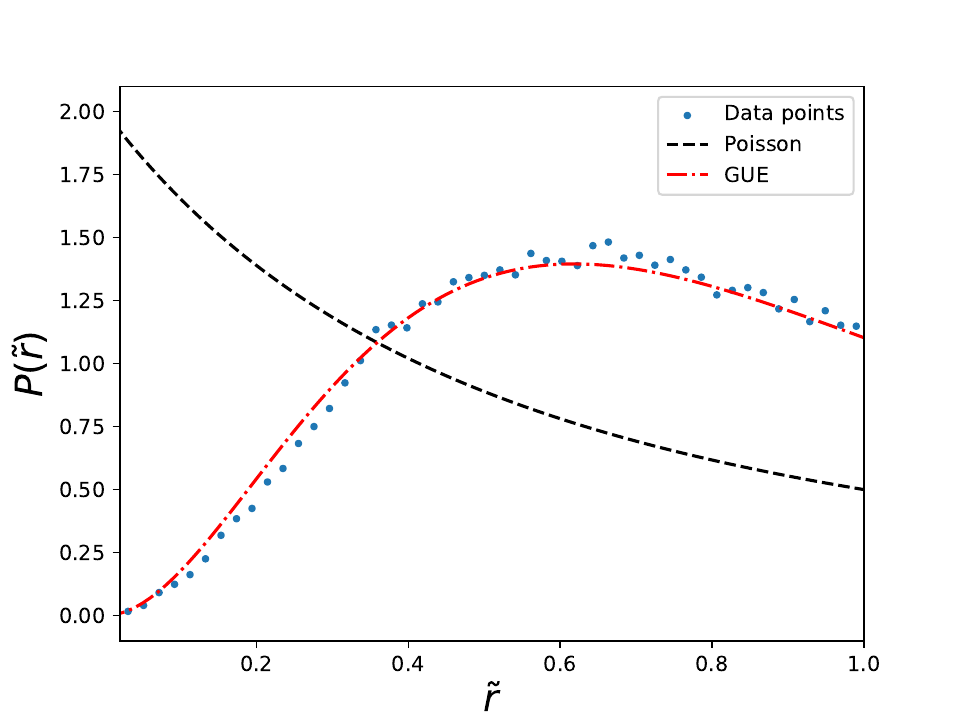}

    \caption{\textbf{Scattering form factor and level statisctics:} Top-panel: Scattering form factor for the imaginary part of non-trivial zeros of Riemann zeta function, without averaging (left) and with averaging (right). We observe that the dip-ramp-plateu structure typical to random-matrix-ensemble only becomes clear after the averaging is done. Bottom-panel: Level statistics for the non-trivial zeros of Riemann zeta function, without averaging (left) and with averaging (right). Again, we notice that it overlaps with GUE (Gaussian Unitary Ensemble) only when the averaging is done.}
    \label{fig:ScFFandLS}
\end{figure}

\noindent Another important issue is the notion of unfolding. By the unfolding procedure it is made sure that the density of eigenvalues after unfolding is almost uniform throughout the data set. It plays an important role in yielding the correct dip-ramp-plateau behavior in the ScFF. It was noted in \cite{Bianchi:2024fsi} that if the data sets of HTT string scattering amplitudes are not unfolded separately, then the ScFF does not capture the 'ramp' behavior. However, even unfolding plays a little role in our scattering complexity calculation, because the plots for scattering complexity in our work \textit{does not use unfolding procedure}, yet we get the clear peak in the complexity which is anyway the key part in identifying chaotic scatterings by our proposal.

\begin{figure}[htbp]
    \centering
    \includegraphics[width = 0.32\linewidth]{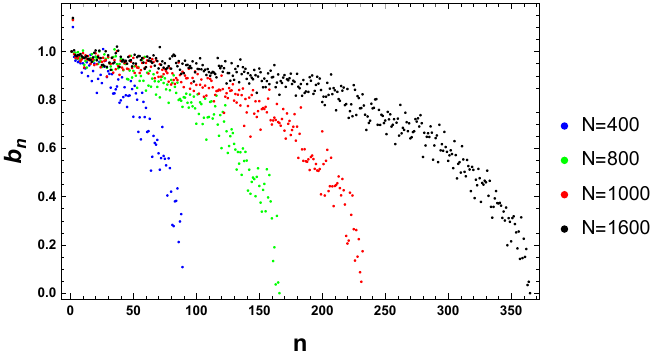}
    \includegraphics[width = 0.32\linewidth]{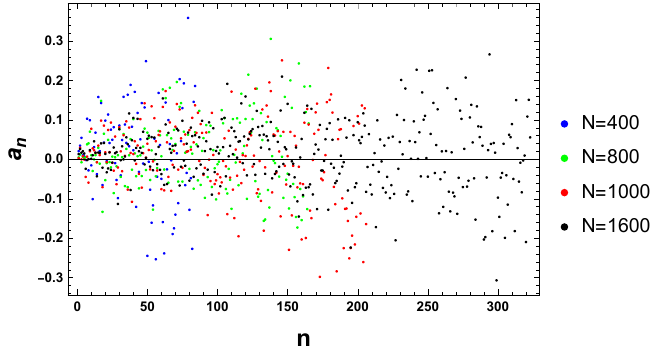}
    \includegraphics[width = 0.32\linewidth]{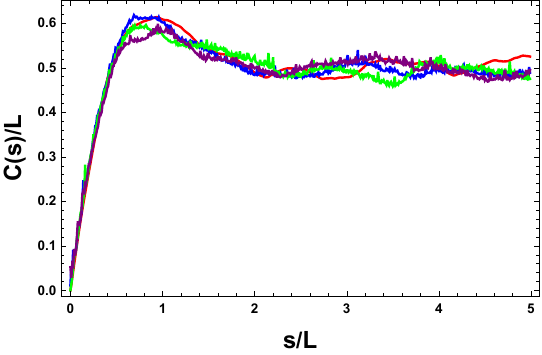}
    \caption{ \textbf{Lanczos coeff. and complexity without averaging:} Left: $b_n$ vs $n$ for single realization of HESS to two tachyons, Middle:$a_n$ vs $n$ for single realization of HESS to two tachyons, Right: Complexity for four single realizatios of HESS to two tachyons (with \(N=800\)). While there are a bit more fluctuations in the single realization plots of complexity as compared to the ones in Fig. \ref{fig:Chaotic}, the presence of peak characterizing chaos is still clearly evident in all the single realizations.}
    \label{fig:singleRealizationplots}
\end{figure}

\noindent \textbf{Chaotic complexity without averaging:} While we discussed the necessity of averaging over many realizations for getting consistent plots of level statistics, and scattering form factor, which is true for the studies of the spectrum of usual Hamiltonians as well, we notice that complexity is less sensitive to the averaging procedure. To elucidate what we mean, we have provided the plots for Lanczos coefficients and the complexity for the HESS scattering to two tachyons case (chaotic) with a single realization in Fig. \ref{fig:singleRealizationplots}. None of the plots look much different from the corresponding averaged plots shown in Figures \ref{fig:anbnintChaotic} and \ref{fig:Chaotic}. To remind the reader, by averaging in case of the HESS to two tachyons case, we mean considering all possible partitions $\{g_n\}$ of $N=\sum_n n g_n$ and $J=\sum_n g_n$. While the main text plot in Fig. \ref{fig:Chaotic} is done by averaging over all possible partitions, in Fig. \ref{fig:singleRealizationplots}, we see that even a single choice of partition shows similar peak and saturation behavior. Hence, to our understanding, complexity appears to be much more robust probe of chaos than the level statistics or the scattering form factor.

\section{Comments on Low Energy Deformation and Truncated Amplitude}\label{sanity checks}

In the main text, we have made a comment that the non-chaotic nature of the Veneziano amplitude is due to the regular spacing of zeros in the logarithmic derivative. Here, we try to see how sensitive this property is to some low-energy modifications as well as to the use of new crossing symmetric representation of tree-level string amplitude.

 i) First, we add a term \(\varepsilon (s_1+s_2)\) to the Veneziano amplitude, with \(\varepsilon\) very small. It is observed that, though the amplitude is deformed significantly for non-zero \(\varepsilon\) (towards small \(z=\cos\theta\)), the spacing of zeros are still almost equal. Hence, while complexity profile can go through minor changes, there is no huge shift from non-chaotic to chaotic due to these low energy modifications.

 ii) Now, we go to the crossing-symmetric series representation of Euler-Beta function. From \cite{Saha:2024qpt} , we use a more general representation,

 \begin{equation}
     \frac{\Gamma (\alpha-s_1)\Gamma(\alpha-s_2)}{\Gamma(\beta - s_1 - s_2)} = \sum_{n=0}^{\infty} \frac{(-1)^{p+1}}{n!}\left(\frac{1}{s_1 - \alpha -n} + \frac{1}{s_2 - \alpha -n} + \frac{1}{\alpha + \lambda+ n}\right) \left(1-\alpha-\lambda + \frac{(s_1+\lambda)(s_2+\lambda)}{\alpha+\lambda+n}\right)_{n+p}
 \end{equation}

 here \(p=2\alpha - \beta\) and \(\lambda > \alpha - \beta\).

 For our purpose, \(\alpha=-1\) and \(\beta=-2\), and we truncate the sum upto \(N_{max}\). For \(N_{max}\gtrsim s_1\), the truncated expression agrees with the actual Beta function, and is reflected in the logarithmic derivative also. But if we truncate \(N_{max}<s_1\), then the amplitude actually deforms away from original Beta function value, still retaining the property of almost equally spaced zeros (no erratic behavior). What this tells us is that the representation in \cite{Saha:2024qpt} for particular choice of $N_{max}>s_1$ works as good as the original Beta function if we compute complexity from the series representation. This is because all we need for our notion of complexity is the structure of zeros of $\mathcal{F}=\frac{d(\log\mathcal A)}{dz} $.

\begin{figure}
    \centering
    \includegraphics[width=0.32\linewidth]{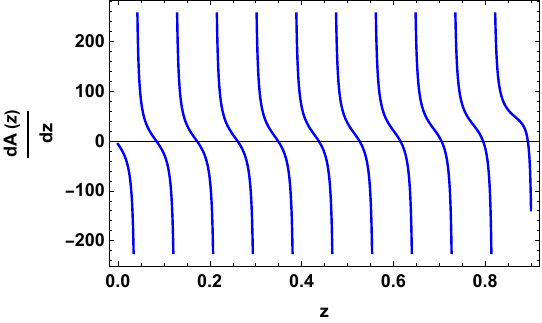}
    \includegraphics[width=0.32\linewidth]{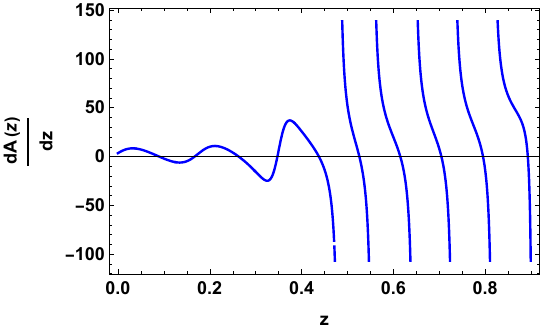}
    \includegraphics[width=0.32\linewidth]{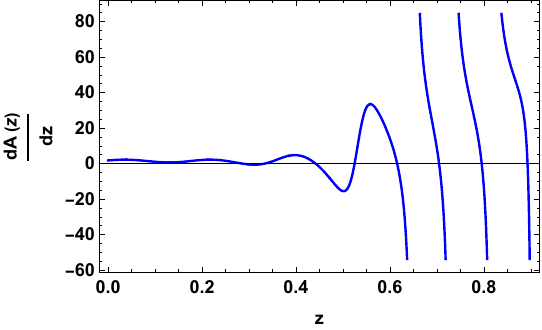}
    \vskip\baselineskip
    \includegraphics[width=0.32\linewidth]{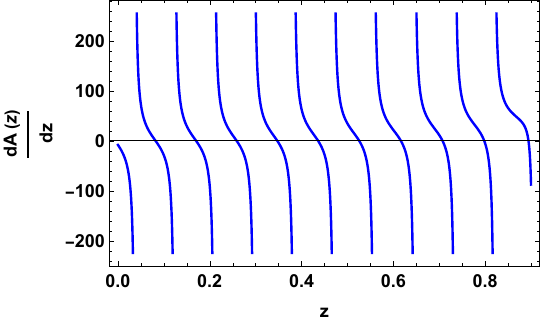}
    \includegraphics[width=0.32\linewidth]{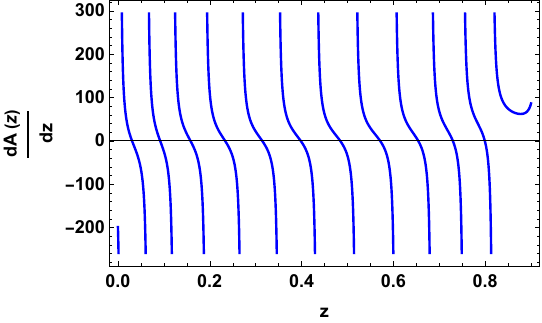}
    \includegraphics[width=0.32\linewidth]{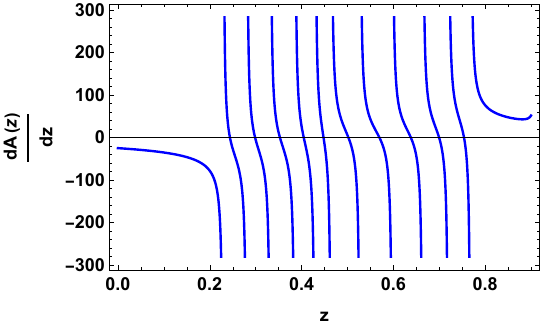}
    
    \caption{Top: Effect of adding low energy contact term \(\varepsilon(s_1+s_2)\) to the Veneziano amplitude, \(\varepsilon=0\) (left), \(\varepsilon=10^{-6}\) (middle) and \(\varepsilon=10^{-5}\) (right) for \(s_1=19.51\). Bottom: Effect of using crossing-symmetric truncated series representation with \(N_{max}\) number of terms, \(N_{max}=20\) (left), \(N_{max}=15\) (middle) and \(N_{max}=10\) (right) for \(s_1 = 19.51\) and \(\lambda = 4.1\). (Here \(A(z) = \log \mathcal A (z)\))}
    \label{fig:lowenergy}
\end{figure}

\bibliographystyle{apsrev4-1}
\bibliography{apssamp}
\end{document}